\documentclass[12pt]{article}
\usepackage[T1]{fontenc}
\usepackage[dvips]{graphicx}
\usepackage[tbtags]{amsmath}
\usepackage{amsopn}
\usepackage{amsfonts}

\title{Abelian duality in three dimensions%
\footnote{supported by KBN grant 5 P03B 072 21}}
\author{Bogus\l aw Broda%
\footnote{e-mail: bobroda@uni.lodz.pl}\,
and
Grzegorz Duniec\\
Department of Theoretical Physics, University of \L \'od\'z\\
Pomorska 149/153, PL-90-236 \L \'od\'z, Poland}

\begin{document}
\baselineskip24pt
\maketitle

\begin{abstract}
Abelian duality on the closed three-dimensional Riemannian manifold
$\mathcal{M}^3$ is discussed. Partition functions for the ordinary $U(1)$ gauge theory and a circle-valued scalar field theory on $\mathcal{M}^3$ are
explicitly calculated and compared. It is shown that the both theories are mutually dual.
\end{abstract}

\noindent
{\bf Keywords:} Abelian duality, three-dimensional Abelian gauge theory, circle-valued scalar free field theory, quantum field theory on a Riemannian manifold

\noindent
{\bf PACS numbers:} 02.40.-k, 11.10.Kk, 11.15.-q, 11.90.+t

\section{Introduction}

At present, Abelian duality is a classical theme in (quantum)
field theory. In four-dimensional case, there are two well-known
papers on electric-magnetic (or $S$-duality) in Abelian gauge
theory, viz.\ [\ref{B01},\ref{B02}]. In three dimensions, one could
cite the other two papers, [\ref{B03}] and [\ref{B04}], but
unfortunately, neither is fully comprehensive nor satisfactory.
The first one ([\ref{B03}]) lacks explicit formulas for the partition
functions and a definite conclusion, whereas the second one ([\ref{B04}]) lacks sufficient generality and
topological aspects are there largely ignored. Moreover, the both
are seemingly contradictory: [\ref{B03}] suggests exact (i.e.\ without any coefficients)   duality
between Abelian gauge field and scalar field, whereas the
corresponding partition functions calculated in [\ref{B04}]
markedly differ.

The aim of this short work is to fill the gap and
clarify some points (the reader is advised to consult the above-mentioned papers for more details which are not repeated here).
In particular, we will explicitly calculate the both partition functions and we will show the mutual duality of the both theories.

\section{Partition function of the Abelian theory}

We consider the connected orientable three-dimensional Riemannian (of
Euclidean signature) manifold $\mathcal{M}^{3}$. The action of $U(1)$ gauge theory on $\mathcal{M}^{3}$ is defined by
\begin{equation}\label{T1}
S[A]= \frac{1}{4 \pi e^{2}}\int_{\mathcal{M}^{3}}F_{A}\wedge \ast F_{A}
=\frac{1}{8\pi e^{2}} \int_{\mathcal{M}^3}d^{3}x \sqrt{g}\,F_{ij}F^{ij},
\qquad
i,j= 1,2,3,
\end{equation}
where $F_{ij}=\partial_{i}A_{j}-\partial_{j}A_{i}$ ($F_{A}=dA$), $e$ is the coupling constant ($e>0$), and $g_{ij}$ is the metric tensor on $\mathcal{M}^{3}$.
A standard form of the partition function is
\begin{equation}\label{T2}
Z_{A}=\int\mathcal{D}A\,e^{- S[A]},
\end{equation}
where $\mathcal{D}A$ denotes a formal integration measure with respect to gauge non-equivalent field configurations, i.e.\ the
Faddeev--Popov procedure is applied.

The partition function consists of several factors:
(1) $Z_{\rm det}$ --- product of (regularized) determinants,
(2) $Z_{\rm class}$ --- sum over the classical saddle points,
(3) $Z_{\rm vol}$ --- the volume of the space of classical minima,
(4) $Z_{0}$ --- a contribution from zero modes.

The form of $Z_{\rm det}$, coming from Gaussian integration with respect to gauge fields and ghosts,
is rather standard, i.e.
\begin{equation}\label{T3}
Z_{\rm det}=\frac{{\det}'\Delta_{0}}
{{\det}'^{1/2}\Delta_1},
\end{equation}
where the prime denotes removal of zero modes and possible regularization of the determinant,
$\Delta_{0}$ is a Laplacian acting on zero-forms
(Faddeev--Popov ghosts), and $\Delta_{1}$ is a Laplacian acting on
one-forms (gauge fields). Obviously, $Z_{\rm det}$ is independent of the
coupling constant $e^2$ and physically corresponds to ``quantum fluctuations''.

The classical part represents the sum over classical saddle
points,
\begin{equation}\label{T4}
Z_{\rm class}^{(A)}=\sum e^{-S[A_{\rm class}]},
\end{equation}
where $A_{\rm class}$ are local minima of the action (1)
corresponding to different line bundles. For non-trivial second
homology of $\mathcal{M}^{3}$ we have the field configurations with
non-zero flux,
\begin{equation}\label{T5}
\int_{\Sigma_{I}}F = 2 \pi m^{I}, \qquad m^{I} \in \mathbb{Z},
\end{equation}
$I= 1,...,b_{2}=\dim H_{2}(\mathcal{M}^{3})$. Then the
solutions of the field equations can be represented by the sum
\begin{equation}\label{T6}
F=2 \pi \sum_{ I } m^{ I } \omega_{ I },
\end{equation}
where $\omega_{I}$ span an integral basis of harmonic two-forms
with normalization
\begin{equation}\label{T7}
\int_{\Sigma_{ I}} \omega_{J}=\delta_{J}^{I}.
\end{equation}
Inserting the expansion \eqref{T6} into the partition function \eqref{T4} we obtain
\begin{equation}\label{T8}
Z_{\rm class}^{(A)}=\sum_{m^{I}} e^{- S[m^{I}]}.
\end{equation}
Here
\begin{equation}\label{T9}
S[m^{I}]=\frac{\pi}{e^{2}} \sum_{I,J} G_{IJ} m^{I} m^{J},
\end{equation}
with
\begin{equation}\label{T10}
G_{IJ}=\frac{1}{2} \int_{\mathcal{M}^{3}} d^{3}x
\sqrt{g}\; \omega_{Iij}\omega_{J}^{ij}.
\end{equation}

The volume of the space of classical minima is determined by a disjoint union of tori (one for each torsion element of ${\rm Tors}(H_1(\mathcal{M}^{3}))$) of dimension
$b_{1}(\mathcal{M}^{3})$ [\ref{B01}], that is,
in our case,
\begin{equation}\label{T11}
Z_{\rm vol}=|{\rm Tors}(H_1)|\cdot(2\pi)^{b_{1}}\cdot{\cal V},
\end{equation}
where the explicit form of $\cal V$ is given by the metric on the torus. Since the metric is determined by the zero modes (which are tangent to the torus [\ref{B01}]), we have (see, Eq.(3.5.5-6) in [\ref{Zu}])
\begin{equation}\label{volmol}
{\cal V}={\det}^{1/2}H_{PR},
\end{equation}
where $H$ is defined by Eq.\eqref{T23}.

Finally, the contribution from zero modes is
\begin{equation}
(2\pi e)^{b_{0}-b_{1}}V^{1/2}=(2\pi e)^{1-b_{1}}V^{1/2},
\end{equation}
where $b_{0}$ corresponds to Faddeev--Popov ghosts, $b_{1}$ is related to Abelian gauge fields, whereas the presence of the volume term $V$ of ${\cal M}^3$ follows from the normalization of the scalar (ghost) zero mode.
Actually, we should divide this by $2\pi$ to obtain an agreement
with ``a direct calculation'' of the path integral (this is
explicitly demonstrated by considering a simple finite-dimensional
example in Appendix of [\ref{B05}]).
Thus,
\begin{equation}\label{T12}
Z_{0}=\frac{1}{2\pi}(2\pi e)^{1-b_{1}}V^{1/2}.
\end{equation}

Collecting all the above terms, i.e. Eq.\eqref{T3}, Eq.\eqref{T8}, Eq.\eqref{T11} and Eq.\eqref{T12}, we obtain an
explicit form of the partition function for the three-dimensional $U(1)$ gauge theory,
\begin{equation}\label{T13}
Z_{A}= \frac{e^{1 - b_1}}{V^{1/2}} |{\rm Tors}(H_1)| {\det}^{1/2}H
\,\frac{{\det}' \Delta_{0} }
{{\det}'^{1/2}\Delta_1} \sum_{m^ {I} } e^{-S[m^{ I }] },
\end{equation}
with $S[m^{I}]$ defined by Eq.\eqref{T9}.

\section{Partition function of the scalar theory}

In this section we will calculate the partition function of the
three-dimensional (single-component) scalar field theory on
$\mathcal{M}^{3}$. In principle, we have the two possibilities: we
could consider an ordinary scalar field assuming values in
$\mathbb{R}$ or a circle-valued scalar field with values in $S$.
A short reflection suggests the second possibility. First of all, for
ordinary scalar field we would not have a chance to reproduce the
reach  topological classical sector of the form \eqref{T8}. Anyway, $\mathbb{R}$-valued theory can be considered as a
topologically trivial sector in the circle-valued one.

The action of the scalar theory is of the form
\begin{equation}\label{T14}
S[\phi]= \frac{e^{2}}{4\pi}\int_{\mathcal{M}^{3}}E_{\phi}\wedge \ast E_{\phi}
=\frac{e^{2}}{4\pi} \int_{\mathcal{M}^{3}} d^{3}x \sqrt{g}\,E_{i} E^{i},
\end{equation}
where $E_{i}=\partial_{i} \phi$ ($E_{\phi}=d \phi$), with $\phi \in [0,2\pi)$. The
rest of the notation is analogous to the one used in the previous
section. The partition function for the scalar theory,
\begin{equation}\label{T15}
Z_{\phi}=\int D\phi \, e^{-S[\phi]},
\end{equation}
can be calculated according to the scheme proposed for $Z_A$.

The determinantal part consists of the only one determinant coming from non-zero modes of the scalar field,
\begin{equation}\label{T16}
Z_{\rm det}={\det}'^{-1/2}\Delta_0.
\end{equation}

As far as the classical part is concerned, we have
\begin{equation}\label{T17}
Z_{\rm class}^{(\phi)}=\sum e^{-S[\phi_{\rm class}]},
\end{equation}
where, analogously to Eq.\eqref{T4}, the sum is taken over the classical saddle points.
But this time, for non-trivial first homology of $\mathcal{M}^{3}$ we have the field configurations
with non-zero circulation,
\begin{equation}\label{T18}
\int_{C_{P}} E = 2\pi n^P, \qquad n^{P} \in \mathbb{Z},
\end{equation}
with $P=1,...,b_{1}={\rm dim}H_{1}(\mathcal{M}^{3})$. Then, the solutions of the field equations can be expressed by the sum
\begin{equation}\label{T19}
E=2\pi\sum_{P} n^{P}\alpha_{P},
\end{equation}
where $\alpha_{P}$  span an integral basis of harmonic
one-forms with the normalization
\begin{equation}\label{onenorm}
\int_{C_P} \alpha_{R}=\delta_{R}^{P}.
\end{equation}
Inserting the expansion \eqref{T19} into the partition function
\eqref{T17} we obtain
\begin{equation}\label{T21}
Z_{\rm class}^{(\phi)}=\sum_{n^{P}} e^{-\widetilde{S}[n^P]},
\end{equation}
where this time
\begin{equation}\label{T22}
\widetilde{S} [n^{P}] = \pi e^{2} \sum_{P,R} H_{PR} n^{P}n^{R},
\end{equation}
with
\begin{equation}\label{T23}
H_{PR}=\int_{\mathcal{M}^{3}} d^{3}x
\sqrt{g} \; \alpha_{Pi}\alpha_{R}^{i}.
\end{equation}

The volume of the space of classical minima, which is the target
space for this model (the circle $S$), is
\begin{equation}\label{T24}
Z_{\rm vol}= 2\pi,
\end{equation}
and the zero-mode contribution (from a single zero mode of the scalar field),
\begin{equation}\label{T25}
Z_{0}=\frac{e}{2\pi}V^{1/2}.
\end{equation}

The final shape of the partition function as the product of
Eq.\eqref{T16}, Eq.\eqref{T21}, Eq.\eqref{T24}, and Eq.\eqref{T25}
assumes the form
\begin{equation}\label{T26}
Z_{\phi}=eV^{1/2}{\det}'^{-1/2}\Delta_0
\sum_{n^{P} } e^{-\widetilde{S}[n^{P}]},
\end{equation}
with $\widetilde{S}[n^{P}]$ defined by Eq.\eqref{T22}.

\section{Duality and final discussion}

To show duality, at least in some limited sense, of the $U(1)$ gauge
theory and the circle-valued scalar field theory, we should perform a Poisson resummation of the classical part of the partition function of the gauge field, Eq.\eqref{T8}, to convert it into an expression proportional to the classical part of the partition function of the scalar field, Eq.\eqref{T17}.

The Poisson resummation of Eq.\eqref{T8} provides
\begin{equation}\label{ZclassA}
Z_{\rm class}^{(A)}={\det}^{-1/2}(e^{-2} G)
\sum_{m^I} \exp[-\pi e^2 \sum_{I,J} G^{IJ}m_I m_J ],
\end{equation}
where $G^{IJ}$ is the inverse matrix to $G_{IJ}$ ($G^{IK}G_{KJ}=\delta_J^I$).
Inserting the identity \eqref{A10} from Appendix,
\begin{equation}\label{QHQ}
G^{IJ}=Q^{IP}H_{PR}Q^{JR},
\end{equation}
into the exponent of Eq.\eqref{ZclassA}, and next using the modular property of the $\theta$-function [\ref{B06}], we get
\begin{equation}\label{ZclassA2}
Z_{\rm class}^{(A)}=e^{b_2}\,{\det}^{-1/2}(G_{IJ})\, {\det}^{-1/2}(Q^{KP})\,
Z_{\rm class}^{(\phi)}
=e^{b_2}\, {\det}^{-1/2} Q_I^P\, Z_{\rm class}^{(\phi)}.
\end{equation}

The presence of determinants of Laplacians signals the appearance of the Reidemeister--Ray--Singer torsion.
In general case, the Reidemeister--Ray--Singer torsion is a product of the two terms (see, Appendix B in [\ref{FW}]): the well-known analytic or determinantal term,
\begin{equation}\label{RRSta}
\tau_a(\mathcal{M}^3)={\det}'^{-3/2} \Delta_0\, {\det}'^{1/2} \Delta_1,
\end{equation}
and a metric (or cohomological) term,
\begin{equation}\label{RRStm}
\tau_m(\mathcal{M}^3)={\det}^{-1/2}G_{(0)}\, {\det}^{1/2}G_{(1)}\, {\det}^{-1/2}G_{(2)}\, {\det}^{1/2}G_{(3)},
\end{equation}
where
\begin{equation}\label{defGs}
G_{(i)}=\int \alpha_{(i)}\wedge \ast \alpha_{(i)},
\end{equation}
with $\alpha_{(i)}$ --- an integral basis of harmonic $i$-forms. In our notation $\alpha_{(0)}=1$, $\alpha_{(1)}=\{\alpha_P\}$, $\alpha_{(2)}=\{\omega_I\}$, $\alpha_{(3)}=V^{-1/2}$, and correspondingly, $G_{(0)}=V$, $G_{(1)}=H_{PR}$, $G_{(2)}=G_{IJ}$, $G_{(3)}=1/V$.
Then
\begin{equation}\label{RRStm2}
\tau_m(\mathcal{M}^3)=\frac{1}{V}{\det}^{1/2}H\, {\det}^{-1/2}G,
\end{equation}
and calculating determinants of the both sides od Eq.\eqref{A11} yields
\begin{equation}\label{RRStm2}
\tau_m(\mathcal{M}^3)=\frac{1}{V}{\det}^{1/2}Q^I_P.
\end{equation}
Now, we obtain from Eq.\eqref{T13} and Eq.\eqref{ZclassA2}
\begin{equation}\label{ZA}
Z_A=e^{1-b_1}\,|{\rm Tors}(H_1)|\, \tau(\mathcal{M}^3)\,{\det}^{1/2}H_{PR}\, {\det}'^{-1/2} \Delta_0\,V^{1/2} e^{b_2}\, {\det}^{1/2} Q_I^P\, Z_{\rm class}^{(\phi)},
\end{equation}
and finally
\begin{equation}\label{Dual}
Z_A=|{\rm Tors}(H_1)|\, \tau(\mathcal{M}^3)\,{\det}^{1/2} Q_{IP}\, Z_{\phi},
\end{equation}
which constitutes our generalized duality relation.
In the case of $b_1=0$, since the Reidemeister--Ray--Singer torsion of the trivial representation is equal to $|{\rm Tors}(H_1)|^{-1}$ (see, Footnote 4 in [\ref{B08}]), and $\det Q=1$, we get a simplification,
\begin{equation}\label{Dual2}
Z_A= Z_{\phi}.
\end{equation}

The authors are very grateful to the referee for his helpful remarks concerning the torsions and the Poisson resummation allowing to avoid several serious errors present in the previous version of the paper.

\section{Appendix}

Let us recall
\begin{equation}\label{A1}
G_{IJ}=\int \omega_I \wedge\ast\omega_J,
\qquad
H_{PR}=\int \alpha_P \wedge\ast\alpha_R,
\end{equation}
where
\begin{equation}\label{A2}
\omega_I \in H^2(\mathcal{M}^3,\mathbb{Z}),
\qquad
\alpha_I \in H^1(\mathcal{M}^3,\mathbb{Z}),
\end{equation}
and
\begin{equation}\label{A3}
\ast\alpha_P=Q_P^I\omega_I,
\qquad
\ast\omega_I=Q_I^P\alpha_P.
\end{equation}
Since
\begin{equation}\label{A5}
\alpha_P=\ast\ast\alpha_P=Q_P^I\ast\omega_I=Q_P^IQ_I^R\alpha_R,
\end{equation}
we have
\begin{equation}\label{A6}
Q_P^IQ_I^R=\delta_P^R.
\end{equation}
Now, from (see, Eq.\eqref{A1} and \eqref{A3})
\begin{equation}\label{A7}
G_{IJ}=\int \omega_I \wedge\ast\omega_J
=Q_J^P\int\omega_I\wedge\alpha_P
\end{equation}
it follows that
\begin{equation}\label{A8}
G_{IJ}Q_R^J=\delta_R^P\int\omega_I\wedge\alpha_P
=\int\omega_I\wedge\alpha_R\equiv Q_{IR}.
\end{equation}
Then
\begin{equation}\label{A9}
G_{IJ}=Q_{IP}Q_J^P,
\end{equation}
and analogously for $H_{PR}$. Consequently, we can manipulate the indices $I$, $J$ with the metric tensor $G_{IJ}$, and the indices $P$, $R$ with $H_{PR}$. Thus, from \eqref{A9} we finally deduce that
\begin{equation}\label{A10}
G^{IJ}=Q^{IP}Q_P^J=Q^{IP}H_{PR}Q^{JR}.
\end{equation}
Let us write down another useful identity,
\begin{equation}\label{A11}
G^{IJ}Q_J^PH_{PR}=Q^I_P.
\end{equation}

\end{document}